\documentclass[floats,floatfix,showpacs,preprint,prd,superscriptaddress,onecolumn,nofootinbib]{revtex4}
\usepackage{graphicx, epsfig, amssymb} %include figure files
\usepackage{amsmath, amsfonts}
\usepackage{bm} %include bold math: \bm{} creates bold letters in math mode
\usepackage{color}
\usepackage[usenames]{xcolor}

\newcommand{\non}{\nonumber}
\newcommand{\be}{\begin{equation}}
\newcommand{\ee}{\end{equation}}
\newcommand{\bea}{\begin{eqnarray}}
\newcommand{\eea}{\end{eqnarray}}

\begin{document}

%%%%%%%

\title{Axionic black branes in the $k$-essence sector of the Horndeski model.}

%%%%%%%

\author{Adolfo Cisterna}
\email{adolfo.cisterna@ucentral.cl} 

\affiliation{Vicerrector\'ia
Acad\'emica, Toesca 1783, Universidad Central de Chile, Santiago,
Chile} 

\affiliation{Dipartimento di Fisica, Universit\`a di Trento,Via
Sommarive 14,\\ 38123 Povo (TN), Italy}

\author{Mokhtar Hassaine}
\email{hassaine@inst-mat.utalca.cl} 

\affiliation{Instituto de
Matem\'{a}tica y F\'{\i}sica, Universidad de Talca, Casilla 747,
Talca, Chile}

\author{Julio Oliva}
\email{juoliva@udec.cl}

\affiliation{Departamento de F\'isica, Universidad de Concepci\'on,
Casilla, 160-C,\\ Concepci\'on, Chile}

\author{Massimiliano Rinaldi}
\email{massimiliano.rinaldi@unitn.it} 

\affiliation{Dipartimento di Fisica, Universit\`a di Trento,Via
Sommarive 14,\\ 38123 Povo (TN), Italy}

\affiliation{TIFPA - INFN, \\ Via Sommarive 14,  38123 Trento, Italy}

\begin{abstract}

We construct new black brane solutions in the context of Horndeski gravity, in particular in its K-essence sector.
These models are supported by axion scalar
fields that depend only on the horizon coordinates. The dynamics of these fields is determined by a K-essence term that includes the standard kinetic term $X$ and a correction of the form $X^{k}$. We find both neutral and charged exact and analytic solutions in $D$-dimensions, which are asymptotically anti de Sitter. Then, we describe in detail the
thermodynamical properties  of the four-dimensional solutions and we compute the dual holographic DC conductivity. 
%%%%%%%%
\noindent

\end{abstract}
%%%%%%

\maketitle

%%%%%%

%%%%%%%%%%%%%%%%%%%%%%%%
\section{Introduction}
%%%%%%%%%%%%%%%%%%%%%%%%

\noindent The observed current Universe is not only expanding but
also accelerating because of the presence of a source to the
Einstein equations that differs from the usual mixture of dark
matter, baryonic matter, and radiation. In fact, the simplest
phenomenological explanation for the acceleration is the presence of
a cosmological constant $\Lambda$ in the Einstein-Hilbert action. At
the quantum level, such a constant can be interpreted as a renormalized vacuum energy. The standard model of cosmology
assumes that the current Universe is dominated by the vacuum energy
together with a large amount of cold dark matter and a tiny fraction
of baryonic matter  and it is called $\Lambda$CMD model.

From a fundamental point of view, however, the cosmological constant
has a series of fundamental and conceptual issues, which makes
alternatives rather appealing \cite{Weinberg:1988cp}. In general
terms, one can replace the cosmological constant with a dynamical
degree of freedom that is often modeled as a fluid with special
properties that goes under the name  of ``dark energy''. In this
way, the matter sector of the theory is implemented by a fluid with
unusual but still reasonable properties, whose dynamics dominates at
late times (for a comprehensive review, see e.g. \cite{amendola}).
Finally, the acceleration of the Universe could also be driven by
the dynamics of the classical counterparts of the standard
model fields \cite{DErinaldi}.

There is a somewhat more radical approach to the problem of dark
energy that relies upon a modification of general relativity (GR) in
the infrared. In other words, this means that the Einstein equations
are different at cosmological scales.  Several models of modified
gravity have been explored during the last decade
\cite{Capozziello:2011et, Clifton:2011jh}. One of the most popular
 is the so-called scalar-tensor theories of
gravity (STT), first proposed in the late sixties by Brans-Dicke
\cite{BD} (for a modern review see e.g. \cite{Faraoni:2004pi}). STT
represent the simplest way to describe a diffeomorphism invariant
theory in four dimensions that avoids Ostrogradski instabilities,
which typically arise in higher order theories \cite{Ostro,
deUrries:1998obu}. The price to pay is to introduce new degrees of
freedom in the form of one or more dynamical scalar fields. Through
a suitable Weyl rescaling of the fields, it is always possible to
write STT in terms of modified gravity actions where the Ricci
scalar $R$ is replaced by some arbitrary function of it, $f(R)$. At
least at the classical level, STT and $f(R)$ gravity are perfectly
equivalent \cite{defelice}.

In STT, gravity is described by the graviton spin-two field and one
or more spin-zero particles, represented by scalar fields. In order
to avoid possible violations of the Einstein equivalence principle the usual prescription is that the scalar fields are only
coupled to the metric  and not to matter particles. This means that,
in the matter action, there is no coupling between the new degrees
of freedom and ordinary matter. However, due to the nonlinearity of
Einstein equations, scalar fields produce a backreaction on the
metric that, in turn,  affects the motion of test particles.
Therefore the dynamics of matter fluids is influenced by scalar fields
even in the case of minimal coupling \cite{Faraoni:2004pi,
Will:2005va}.

The most general STT constructed in four dimensions and yielding at
most second order equations of motion is known as the Horndeski model
\cite{Horndeski1}, which is better known in its modern version as
the Galileon theory \cite{Nicolis}. The latter is a STT coming from the
generalization of the decoupling limit of the brane-inspired
Dvali-Gabadadze-Porrati model \cite{Dvali}. Galileon theory,
which exhibits Galilean symmetry in Minkowski spacetime, was further
covariantized in \cite{Esposito} and it was finally shown to be
equivalent to the original Horndeski action; see \cite{Koba}.

The Galileon action exhibits shift symmetry and, in its  covariant
form, is given by
\begin{multline}
\mathcal{L}=\mathcal{K}(X)-G_{3}(X)\square\phi+G_{4}(X)R+G_{4,X}(X)[(\square\phi)^2-(\nabla_{\mu}\nabla_{\nu}\phi)^{2}]+G_{5}(X)G_{\mu\nu}\nabla^{\mu}\nabla^{\nu}\phi
\\
-\frac{G_{5,X}}{6}[(\square\phi)^3+2(\nabla_{\mu}\nabla_{\nu}\phi)^3-3\square\phi(\nabla_{\mu}\nabla_{\nu}\phi)^2].
\label{lagran}
\end{multline}
Here, $X$ represents the canonical kinetic term for the scalar field
$\phi$, while $\mathcal{K}$ and $G_{i}$ are arbitrary functions of
$X$. Each function can be generalized to the case in which it also
explicitly depends on the scalar field itself. Nevertheless, in such
a case, the shift invariance of theory is lost and this considerably
complicates the integration of the field equations.

Many sectors of the theory  \eqref{lagran}  have been investigated
in cosmology. For instance, it was shown that (\ref{lagran})
contains a subset which possesses a self-tuning mechanism that
allows to circumvent Weinberg's theorem on the cosmological constant
\cite{Charmousis:2011bf}. Moreover, the sector defined by the
nonminimal kinetic coupling controlled by the Einstein tensor
exhibits interesting inflationary properties without the need of
\emph{ad hoc} potential terms \cite{Germani:2010gm, Amendola:1993uh,
Sushkov:2009hk, Myrzakulov:2015ysa, namur}. Likewise, the nonminimal
coupling between the Einstein tensor and the scalar field kinetic
term can, on large scales, mimic cold dark matter and also flatten
the rotational curves of galaxies \cite{maxPDE}. Finally, several
works have been devoted to the study of cosmological perturbations
with the aim of finding observable deviations from GR in large-scale
structures and the conditions on the parameter space that avoid too
large gravitational instabilities \cite{pert}.

Technically speaking, the so-called $k$-essence models of dark energy
belong to the class of gravitational theories represented by
\eqref{lagran}. In $k$-essence, the acceleration of the Universe (both
at early and late times) can be driven by the kinetic energy instead
of the potential energy of the scalar field
\cite{ArmendarizPicon:2000ah}. The model was first introduced in
\cite{ArmendarizPicon:1999rj} and then specifically used as dark
energy models in \cite{Chiba:1999ka, Garriga:1999vw, Rendall:2005fv,
Tsujikawa:2010sc, Myrzakul:2016ets, Chimento:2003zf}. These models
are characterized by a nonlinear kinetic term for the scalar field
and are expressed in (\ref{lagran}) by the arbitrary function
$\mathcal{K}(X)$ (together with with $G_{4}=1$ and $G_{3}=G_{5}=0$). Quantum and classical stability of $k$-essence have
been investigated  \cite{Babichev:2007dw}. In particular, the
classical stability and their perturbations are crucial to
discriminate the model from standard GR in view of the forthcoming
Euclid mission \cite{euclidreview}.

One fundamental step that may put the theory on a solid theoretical
foot is the construction of black hole solutions. In principle,
there is a no-hair theorem that prevents the existence of
nontrivial black hole solutions in Galileon gravity
\cite{Hui:2012qt}. However, there are ways to get around this theorem and several
black hole solutions have been found for particular sectors of
(\ref{lagran}), in particular, the one containing the nonminimal
coupling between the Einstein tensor and the kinetic term. Spherically symmetric solutions were found in
\cite{rinaldi, adolfo2, Minam1, minas} where their thermodynamical
properties were also studied.\footnote{This model has been also
thoroughly investigated in the context of astrophysical configurations
such as neutron and boson stars \cite{Cisterna:2015yla,
Cisterna:2016vdx, Maselli:2016gxk, Brihaye:2016lin}.} Moreover,
anti-de Sitter (AdS), asymptotically flat stealth and Lifshitz
solutions with a self-tuned effective cosmological constant were
found making use of a time-dependent scalar field in
\cite{charmousis1,Bravo-Gaete:2013dca}. Charged solutions were found
in \cite{adolfo3, Babichev:2015rva}. Recently also, for the sectors
of \eqref{lagran} controlled by $G_3$ and $G_4$, analytical and
numerical solutions have been found \cite{Babichev:2016fbg,
Babichev:2017guv}.\footnote{In \cite{Charmousis:2015txa} solutions have been obtained considering a kinetic coupling controlled by the Gauss-Bonet invariant.}

There is still  one  sector of \eqref{lagran}, where black holes
solutions are little known, that is the $k$-essence sector governed by
${\cal K}(X)$. This work aims to fill this gap, at least partially,
by exploring black hole configurations in the sector of
\eqref{lagran} that contains terms like $X^k$, in addition to the
usual kinetic term.\footnote{These kind of Lagrangians are used to obtain cosmological models with equation of state parameter satisfying $\omega<1$ \cite{Melchiorri:2002ux}.} To construct our solutions, instead of
considering a spherically symmetric scalar field, we use axion
fields which depend linearly on the Cartesian coordinates along the flat horizon. We see later that these kinds of configurations are used in the context of dual condensed matter systems due to the fact that they break the translational invariance of the dual field theory \cite{Andrade:2013gsa}. This
is an easy way to circumvent the no-hair theorem \cite{Hui:2012qt},
which is mostly based on the fact that the equation of motion for
the scalar field derived from (\ref{lagran}) is given by a current
conservation law of the form $\nabla_{\mu}J^{\mu}=0$. For the case
of spherically symmetric scalar fields, this current is given by the
component $J^r$ whose modulus diverges at the horizon.  In the case
studied here,  our axion fields yield a finite current on the black
hole horizon while simultaneously satisfying the Klein-Gordon
equation. Moreover, the contribution to the equations of motion
coming from the K-essence term is still spherically symmetric and
the energy of the solutions remains finite.\footnote{This is similar
to what happens in the case of the linearly time-dependent scalar
fields considered in \cite{charmousis1}.}

In Ref. \cite{Bardoux:2012aw}, a static black brane with axionic
charge generated by the presence of two $3$-form fields was
presented. The symmetry of the solution is endowed with a planar horizon with a lapse function mimicking that of a hyperbolic black hole in AdS. This apparent discrepancy between the horizon topology and the
metric behavior was shown to be due to the presence of the axionic
charges, which  play the role of an effective curvature
term. The thermodynamical properties and the  possibility of phase transitions
 were also reported in \cite{Bardoux:2012aw}. These
ideas were also applied to construct black branes with a
source given by a scalar field nonminimally coupled to gravity
\cite{Bardoux:2012tr, Caldarelli:2013gqa}.\footnote{A new planar
solution with a conformally coupled scalar field will be presented
in \cite{ADOLFO} This solution represents a novel generalization of
the Bekenstein black hole plus cosmological constant, without self
interaction and free of self-tuned parameters.} Planar/toroidal
black holes with a scalar field are of special interest in the
context of the AdS/CFT correspondence \cite{Maldacena:1997re} due, in particular, to their
applications in nonconventional superconductor systems
\cite{Hartnoll:2009sz, Horowitz:2010gk}. Within this approach, the
nonzero condensate behavior of the unconventional superconductors
can be reproduced by means of a hairy black hole at low temperature
with a hair that should disappear as the temperature increases.
Usually, planar/toroidal solutions suffer from singular behaviors
due to the lack of a curvature scale on the horizon. Nevertheless,
this situation is successfully circumvented using axion fields which
are homogeneously distributed along the horizon coordinates,
providing in this way an effective curvature scale which makes the
spacetime nonsingular. Several solutions with these ingredients
have been reported in order to study different aspects of their
holographic dual systems \cite{Andrade:2013gsa, Davison:2014lua,
Gouteraux:2014hca, Baggioli:2015zoa, Alberte:2016xja, Baggioli:2014roa, Caldarelli:2016nni, Kuang:2017cgt}. A very interesting application is the construction of homogeneous black string and black p-branes with negative cosmological constant, with no more ingredients that minimally coupled scalar fields \cite{Cisterna:2017qrb}. Moreover,
recently these ideas have been applied to the case of Horndeski
theory, specifically to the nonminimal kinetic coupled sector
\cite{Jiang:2017imk, Feng:2017jub, Baggioli:2017ojd}.\\
The paper is organized as follows: Section II is devoted to the description of
our model, its principal properties and the equations of motion. In
Sec. III we construct asymptotically AdS black brane solutions,
including the case where electric and magnetic monopole charges are
considered. The particular case of vanishing cosmological constant
is also studied. Section IV is devoted to the thermodynamical
analysis of the AdS solutions while in Sec. V we present some
holographic applications. Finally in Sec. VI we give some final
remarks and outline some possible extensions. In the appendix, we report the higher-dimensional extension of the solution.

%%%%%%%%%%%%%%%%%%%%%%%
\section{The model}
%%%%%%%%%%%%%%%%%%%%%%%
\noindent We consider the following K-essence Lagrangian in four
dimensions
\begin{equation}
\mathcal{L}=\mathcal{K}(X_1,X_2)
\end{equation}
where the two scalar fields, with their kinetic terms $X_{1}$ and
$X_2$, correspond to the two axion fields. As considered below, the
axion fields are homogenously distributed along the coordinates
of the planar horizon. This explains the reasons for
considering two axion fields in four dimensions. As mentioned
before, we study the case in which the dynamics of each scalar
field is governed by a standard kinetic term plus a nonlinear
contribution given by an arbitrary power of $X$. More precisely, we
consider a $\mathcal{K}$-term of the form
$\mathcal{K}(X_i)=-\sum_{i=1}^{2}(X_i+\gamma X_i^k)$, and hence our
four-dimensional action reads
\begin{equation}
\mathcal{I}[g_{\mu\nu},\phi_i]=
\int\left[\kappa(R-2\Lambda)-\sum_{i=1}^{2}\left(\frac12
\nabla^{\mu}\phi_{i}\nabla_{\mu}\phi_{i}+\gamma \left(\frac12
\nabla^{\mu}\phi_{i}\nabla_{\mu}\phi_{i}\right)^k\right)\right]d^4x\sqrt{-g}\,,
\label{action}
\end{equation}
where we have defined  $X_{i}=\frac12
\nabla^{\mu}\phi_{i}\nabla_{\mu}\phi_{i}$ with $i=1,2$. The coupling
$\gamma$ (with mass dimension $4-4k$)is supposed to be positive in order to avoid phantom
contributions.\footnote{Recently, solutions for the minimally
coupled case with phantom axion fields where studied in
\cite{Zhang:2017tbf}.} For $\gamma=0$ we recover the case of two
minimally coupled scalar fields studied in \cite{Bardoux:2012aw,
Andrade:2013gsa}. Even if the solutions can be constructed in
arbitrary dimensions we focus our attention on the
four-dimensional case, leaving the D-dimensional extension to
Appendix A. The variation of the action with respect to the metric
yields the following Einstein equations,
\begin{equation}
\kappa(G_{\mu\nu}+\Lambda
g_{\mu\nu})=\frac{1}{2}\sum_i\left[\partial_{\mu}\phi_i\partial_{\nu}\phi_i-g_{\mu\nu}X_i+\gamma(kX_i^{k-1}\partial_{\mu}\phi_i\partial_{\nu}\phi_i-g_{\mu\nu}X_i^k)\right]\,,
\label{EOM}
\end{equation}
while the Klein-Gordon equation takes the form
\begin{equation}
\left[(1+\gamma kX_i^{k-1})g^{\mu\nu}+\gamma
k(k-1)X_i^{k-2}\nabla^{\mu}\phi_i\nabla^{\nu}\phi_i\right]\nabla_{\mu}\nabla_{\nu}\phi_i=0.
\label{KG}
\end{equation}
We now impose the planar metric ansatz
\begin{equation}
ds^2=-F(r)dt^2+\frac{dr^2}{G(r)}+r^2(dx_1^2+dx_2^2)\,,
\end{equation}
and we assume that the axion fields depend on the coordinates
$(x_1,x_2)$ only.\footnote{In \cite{Bardoux:2012aw} the authors considered Einstein gravity with a source given by two 3-form fields (whose Hodge duals can be identified with the exterior derivatives of two scalar fields). A Birkoff's like theorem was established where it is shown that each of the two 3-form fields must depend on one for the transverse spatial coordinate.} In the case $F=G$, the
Klein-Gordon equations are easily solved by
\begin{equation}
\phi_1=\lambda x_1,\ \phi_2=\lambda x_2\ .  \label{axions}
\end{equation}
Note that these scalar fields can be dualized to construct solutions with backreacting 2-forms, $B_{(2)}$, by setting
\begin{equation}
H_{(3)}^{(i)}=dB^{(i)}_{(2)}=\star d\phi_i,\ \ i=1,2\ ,
\end{equation}
where $\star$ denotes the Hodge dual.

In order to ensure that the solutions of the previous equations do
not generate ghosts, it is important to check whether or not they
satisfy the null energy condition given by
\begin{equation}
T_{\mu\nu}n^{\mu}n^{\nu}\geq0\,,\quad i=1,2\,,  \label{null}\,.
\end{equation}
Classical stability is instead guaranteed by a positive sound speed, namely
\begin{equation}
c_s^2=\frac{\mathcal{K},_{X_i}}{{\mathcal{K},_{X_i}+2X_i}\mathcal{K},_{X_i,X_i}}>0. \label{sound}
\end{equation}
In our model a sufficient condition to satisfy simultaneously both requirements is  $k>1/2$. As we discuss below,  the further restriction $k>3/2$ also guarantees that the solutions asymptotically match the GR ones and have finite ADM mass.

%%%%%%%%%%%%%%%%%%%%%%%%%%%%%%%%%%%%%%%%%%%%%%%
\section{K-essence Black Holes with Axions}
%%%%%%%%%%%%%%%%%%%%%%%%%%%%%%%%%%%%%%%%%%%%%%%

\noindent Under the conditions described above Eqs.\ (\ref{EOM}) and (\ref{KG}) have the following exact black brane
solution:
\bea\label{sol}
F(r)&=&G(r)=\frac{r^2}{l^2}-\frac{2M}{r}-\frac{\lambda^2}{2\kappa}+\frac{\gamma\lambda^{2k}}{2^k(2k-3)\kappa}r^{2(1-k)}\,\\\non\\\non
\phi_1&=&\lambda x_1,\qquad \phi_2=\lambda x_2\,.
\eea
The case $k=3/2$
needs to be integrated separately and it yields a
logarithmic branch as reported in Appendix A. Looking at this four-dimensional solution we observe that for $1/2<k<3/2$, the asymptotic behavior of our AdS solutions differs
from the standard ones defined in \cite{Henneaux:1985tv}, and, as a
consequence, configurations with infinite mass could be obtained. Thus, from now on we consider only the case $k>3/2$, which allows the use of
standard methods to compute the mass of our solutions.

It is evident that the effect of the axion fields is to
include an effective hyperbolic curvature scale on the metric
proportional to the axion parameter $\lambda$. By setting
$\gamma=0$, we find the solution first described in
\cite{Bardoux:2012aw}.

These solutions can be easily generalized to include electric and
magnetic monopole charges. In order to do this it is sufficient to
include in the action (\ref{action}) the standard Maxwell term
\begin{equation}
S=-\frac{1}{4}\int{F_{\mu\nu}F^{\mu\nu}}d^4x\sqrt{-g}.
\end{equation}
Then, the Maxwell equation
\begin{equation}
\nabla_{\mu}F^{\mu\nu}=0\,,
\end{equation}
is easily solved  by
\begin{equation}
A=-\frac{Q_e}{r}dt+\frac{Q_m}{2}(x_1dx_2-x_2dx_1)\,,
\label{potential}
\end{equation}
where $Q_e$ and $Q_{m}$ are the electric and magnetic monopole
charges. Finally, the general charged solution of the Einstein equations reads
\begin{equation}
F(r)=G(r)=\frac{r^2}{l^2}-\frac{2M}{r}-\frac{\lambda^2}{2\kappa}+\gamma\frac{\lambda^{2k}}{2^k(2k-3)\kappa}r^{2(1-k)}+\frac{1}{4\kappa
r^2}(Q_e^2+Q_m^2). \label{metriccharged}
\end{equation}
Solutions (\ref{sol}) and (\ref{metriccharged}) are the neutral and
charged K-essence generalization of the solutions found in
\cite{Bardoux:2012aw, Andrade:2013gsa}, which are known to possess
interesting holographic properties. We will discuss a particular 
application in Sec. V.

A very interesting case is the one corresponding to $\Lambda=0$. In particular, the uncharged solution takes
the form
\begin{equation}
F(r)=G(r)=-\frac{2M}{r}-\frac{\lambda^2}{2\kappa}+\gamma\frac{\lambda^{2k}}{2^k(2k-3)\kappa}r^{2(1-k)}.
\end{equation}
This solution can have two horizons. To show this in a simple way let us consider the case $k=2$. Then, the horizon location can be found algebraically by solving the equation
\begin{equation}
F(r)=G(r)=-\frac{2M}{r}-\frac{\lambda^2}{2\kappa}+\gamma\frac{\lambda^{4}}{4\kappa
r^2}.
\end{equation}
The two distinct solutions are
\begin{eqnarray}
r_1=-\frac{2M\kappa+\sqrt{4M^2\kappa^2+2\lambda^6\gamma}}{2\lambda^2}\,,\\
r_2=\frac{-2M\kappa+\sqrt{4M^2\kappa^2+2\lambda^6\gamma}}{2\lambda^2}\,.
\end{eqnarray}
For $\gamma>0$ there is just one positive root that corresponds to a cosmological horizon ($r=r_c=r_2$) which surrounds a curvature singularity located at the horizon. However, if both $\gamma$ and $M$ are negative, it is possible to find two horizons. This is evident upon the substitutions $M\rightarrow -|M|$ and $\gamma\rightarrow
-|\gamma|$, which gives the location of an event and a cosmological horizon located respectively at $r=r_h$ and $r=r_c$, with
\begin{eqnarray}
r_c=\frac{2|M|\kappa+\sqrt{4|M|^2\kappa^2-2\lambda^6|\gamma|}}{2\lambda^2}\\
r_h=\frac{2|M|\kappa-\sqrt{4|M|^2\kappa^2-2\lambda^6|\gamma|}}{2\lambda^2}.
\end{eqnarray}
As we mentioned above, negative values of  $\gamma$ could induce violations of
the null energy condition or nonhyperbolicity of the Klein-Gordon
equation. However, this violation may be hidden behind the
event horizon provided the condition
\begin{equation}
|M|>\frac{7\sqrt{3}\lambda^3}{12\kappa}\sqrt{|\gamma|}
\end{equation}
is satisfied.

%%%%%%%%%%%%%%%%%%%%%%%%%%%%%%%%%%%%%%%%%%%%%%%%%%%%%%%%%%%%%%%%%%%
\section{Thermodynamical properties of AdS K-essence black branes}
%%%%%%%%%%%%%%%%%%%%%%%%%%%%%%%%%%%%%%%%%%%%%%%%%%%%%%%%%%%%%%%%%%%
\noindent In order to explore some holographic applications, we first provide a complete and detailed
analysis of the thermodynamic features of the electrically charged
AdS solutions. Note that such studies have been done for
Horndeski black holes with sources given by scalar fields, see e.g.
\cite{Bravo-Gaete:2014haa}.

In our case, the thermodynamics analysis is carried out through
the Euclidean approach. In this case, the partition function for a
thermodynamical ensemble is identified with the Euclidean path
integral in the saddle point approximation around the classical
Euclidean solution \cite{Gibbons:1976ue}. Since we are interested in a
static metric with a planar base manifold, it is enough to consider
the following class of metric,
$$
ds^2=N(r)^2F(r)d\tau^2+\frac{dr^2}{F(r)}+r^2\left(dx_1^2+dx_2^2\right),
$$
where $\tau$ is the periodic Euclidean time related to the
Lorentzian time by $\tau=i\,t$, and the radial coordinate $r_{h}\leq r<\infty$. Now, in order to have a
well-defined reduced action principle with a Euclidean action
depending only on the radial coordinate, some precautions must be
taken. Indeed, in the present case, we are interested in
configurations where the scalar fields $\phi_i$ do not depend on the
radial coordinate but rather on the planar coordinates.
Nevertheless, since the scalar fields only appear in the action
through their derivatives that are constants, we can ``artificially''
introduce radial scalar fields and their associated ``conjugate
momentum'' as
\begin{eqnarray}
&&\Psi_i(r):=\int_0^r \partial_i\phi_i\,dr,\qquad
\Pi_{(i)}:=-\frac{1}{2}\partial_r\Psi_i(r),\qquad
\hat{\Psi}_i(r):=\int_0^r N\,\partial_r\Psi_i \,dr,\\
&&\Psi_{i,k}(r):=\int_0^r (\partial_i\phi_i)^k\,dr,\quad
\Pi_{(i,r)}:=-\frac{1}{2}\partial_r\Psi_{i,k},\quad
\hat{\Psi}_{i,k}(r):=\int_0^r\frac{N}{2^{k-1}}r^{2(1-k)}\partial_{r}\Psi_{i,k}
dr.\nonumber
\end{eqnarray}
Under this prescription, the Euclidean action $I_E$ is given by
\begin{eqnarray}
I_E=\sigma\beta \int_{r_h}^R\sum_{i=1}^2
&&\Big\{N\Big[2\Pi_{(i)}^2+\frac{2\gamma}{2^{k-1}}r^{2(1-k)}\Pi_{(i,k)}^2+\frac{1}{2r^2}\Pi_A^2+2\kappa
r F^{\prime}+2\kappa F-\frac{6\kappa
r^2}{l^2}\Big]\nonumber\\
&&-2\hat{\Psi}_i\Pi_{(i)}^{\prime}-2\gamma
\hat{\Psi}_{i,k}\Pi_{(i,k)}^{\prime}-A\Pi_A^{\prime}\Big\}dr+B_E,
\end{eqnarray}
where $\beta$ is the inverse of the temperature, $\sigma$ stands for
the volume of the two-dimensional compact flat space and $\Pi_{A}$
denotes the conjugate momentum to the vector potential $A$,
$$
\Pi_{A}=-\frac{r^2A^{\prime}}{N}.
$$
The Euclidean action is obtained in the limit $R\to\infty$ and the boundary term
$B_E$ is  fixed by requiring that the 
action has a well-defined extremum, i.e. $\delta I_E=0$. It is easy to check that the field equations obtained by varying the
reduced action yield  the electrically charged AdS solution
(\ref{metriccharged}) with $Q_m=0$. In fact the variations with
respect to $F$ and $A$ give respectively $2\kappa r N^{\prime}=0$
and $\Pi_A^{\prime}=0$. The first equation implies that
$N$ is constant and without loss of generality, can
be taken to be $N=1$. The second equation imposes the electric
potential to have the Coulomb form $A_t=\frac{Q_e}{r}$. On the
other hand, the variations with respect to the conjugate momenta
$\Pi_{(i)}$, $\Pi_{(i,k)}$ and $\Pi_A$ yield equations that are trivially satisfied while
those obtained by variation with respect to $\hat{\Psi}_i$ and $\hat{\Psi}_{i,k}$ can be easily
solved by choosing
$$
\Pi_{(i)}=-\frac{1}{2}\lambda,\quad\Pi_{(i,k)}=-\frac{1}{2}\lambda^k
\Rightarrow \phi_i=\lambda x_i.
$$
Finally, the equation obtained by varying $N$
\begin{eqnarray*}
2\Pi_{(i)}^2+\frac{2\gamma}{2^{k-1}}r^{2(1-k)}\Pi_{(i,k)}^2+\frac{1}{2r^2}\Pi_A^2+2\kappa
r F^{\prime}+2\kappa F-\frac{6\kappa r^2}{l^2}=0,
\end{eqnarray*}
gives rise to a differential equation for the metric function $F$
whose integration yields (\ref{metriccharged}).

Now, in order to compute the boundary term, we consider the
formalism of the grand canonical ensemble where the temperature
$\beta^{-1}$ as well as the ``potentials" at the horizon $A(r_h),
\hat{\Psi}_i(r_h)$ and $\hat{\Psi}_{i,k}(r_h)$ are fixed. The
extremal condition $\delta I_E=0$ implies that the contribution of
the boundary term must be given by
$$
\delta B_E=\Big[\sum_{i=1}^2\left(-2\kappa\sigma \beta N r \delta
F+2\sigma\beta\hat{\Psi}_i\delta\Pi_{(i)}+2\sigma\beta\gamma
\hat{\Psi}_{i,k}\delta\Pi_{(i,k)}+\sigma\beta
A\delta\Pi_A\right)\Big]_{r=r_h}^{r=R}.
$$
Without loss of generality, we can set again $N=1$, and the
contribution at infinity reduces to
$$
\delta B_E(R)=4\kappa\sigma\beta\delta M\Longrightarrow
B_E(R)=4\kappa\sigma\beta M.
$$
At the horizon, in order to avoid the conical singularities, the
variation of the metric function at the horizon is given by $\delta
F\vert_{r_h}=-\frac{4\pi}{\beta}\delta r_h$, and hence one gets
\begin{eqnarray*}
B_E(r_h)=4\pi\kappa\sigma r_h^2+\sigma\beta A(r_h)
Q_e-\sigma\beta\sum_i\left(\lambda
\hat{\Psi}_i(r_h)+\gamma\lambda^k\hat{\Psi}_{i,k}(r_h)\right).
\end{eqnarray*}
Finally, the boundary term becomes
\begin{equation}
B_E=4\kappa\sigma\beta M-4\pi\kappa\sigma r_h^2-\sigma\beta A(r_h)
Q_e+\sigma\beta\sum_i\left(\lambda
\hat{\Psi}_i(r_h)+\gamma\lambda^k\hat{\Psi}_{i,k}(r_h)\right).
\end{equation}
\normalsize The Euclidean action is related to the Gibbs free energy
${\cal G}$ by
$$
I_E=\beta{\cal G}=\beta{\cal M}-{\cal S}-\beta A(r_h){\cal
Q}_e-\beta\sum_i\left(\hat{\Psi}_i{\cal Q}_i+\hat{\Psi}_{i,k}{\cal
Q}_{i,k}\right).
$$
The mass ${\cal M}$ is given by
\begin{eqnarray}\small
{\cal M}&=&\left(\frac{\partial
I_E}{\partial\beta}\right)_{\hat{\Psi}_i,\hat{\Psi}_{i,k}}-\frac{\hat{\Psi}_i}{\beta}\left(\frac{\partial
I_E}{\partial\hat{\Psi}_i}\right)_{\beta}-\frac{\hat{\Psi}_{i,k}}{\beta}\left(\frac{\partial
I_E}{\partial\hat{\Psi}_{i,k}}\right)_{\beta}=4\kappa\sigma
M\nonumber\\
&=&4\kappa\sigma \left(\frac{r_h^3}{2l^2}-\frac{\lambda^2
r_h}{4\kappa}+\frac{\gamma\lambda^{2k}r_h^{3-2k}}{2^{k+1}(2k-3)\kappa}+\frac{Q_e^2}{8\kappa
r_h}\right),
\end{eqnarray}
while the entropy ${\cal S}$, the electric charge ${\cal Q}_e$ and
the axion charges ${\cal Q}_i, {\cal Q}_{i,k}$ are defined by
\begin{eqnarray}
{\cal S} &=& \beta \left(\frac{\partial
I_E}{\partial\beta}\right)_{\hat{\Psi}_i,\hat{\Psi}_{i,k}}-I_E=4\pi\kappa\sigma
r_h^2,\qquad {\cal Q}_{e}= -\frac{1}{\beta}\left(\frac{\partial
I_E}{\partial
A(r_h)}\right)_{\beta}=\sigma Q_e,\\
{\cal Q}_{i} &=& -\frac{1}{\beta}\left(\frac{\partial
I_E}{\partial\hat{\Psi}_i}\right)_{\beta}=-\sigma\lambda, \qquad
{\cal Q}_{i,k}= -\frac{1}{\beta}\left(\frac{\partial
I_E}{\partial\hat{\Psi}_{i,k}}\right)_{\beta}=-\sigma\gamma\lambda^k.
\nonumber
\end{eqnarray}
With these results it is trivial to see that the first law holds, namely
$$
d{\cal M}=T d{\cal S}+A(r_h)d{\cal
Q}_e+\sum_i\left(\hat{\Psi}_i(r_h) d{\cal Q}_i+\hat{\Psi}_{i,k}(r_h)
d{\cal Q}_{i,k}\right).
$$
We conclude this section by comparing our results with those
obtained recently for a similar model with phantom axion fields
\cite{Zhang:2017tbf}. In this reference, the thermodynamics analysis
of the phantom black hole solution is carried out without
considering the axion parameter constant $\lambda$ as an axion
charge. Because of that, the thermal properties of the phantom
solution are quite analogous to those of the Schwarzschild-AdS black
hole. The importance of considering axion
parameter constant $\lambda$ as an axion charge is particularly important when holographic applications and phase transitions are studied.

%%%%%%%%%%%%%%%%%%%%%%%%%%%%%%%%%%%%%
\section{Holographic DC conductivity}
\begin{figure}
\setlength\unitlength{1mm}
\includegraphics[width=170mm]{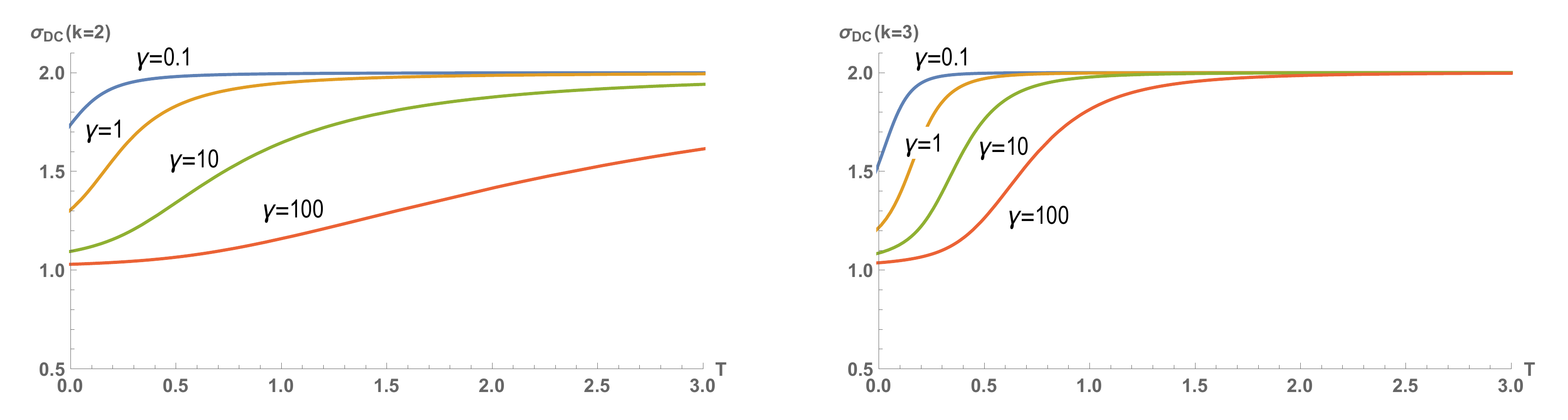}
\caption[FIG:A]{DC conductivities as a function of temperature for the cases $k=2$ (left panel) and $k=3$ (right panel). Different curves on each panel correspond to different values of the derivative self-interaction coupling $\gamma$. We have set the chemical potential $\mu$=1, as well as the AdS radius $l=1$, the axions constant $\lambda=1$ and $\kappa=1$.} 
\label{fig.0}
\end{figure}

%%%%%%%%%%%%%%%%%%%%%%%%%%%%%%%%%%%%%
Charged black brane solutions provide a perfect setup to compute
holographic conductivities \footnote{See also \cite{Thurs} for the computation of thermoelectric transport coefficients of systems which are dual to five-dimensional, charged black holes with horizons modeled by Thurston geometries.} \cite{Andrade:2013gsa,Jiang:2017imk, Blake:2013bqa,
Iqbal:2008by, Donos:2014cya, Banks:2015wha}. This can be done by
constructing a conserved current with radial dependence from which
it is possible to obtain the holographic properties on the boundary
in terms of the black hole horizon data. Here, we are interested in the
effects of the nonlinear kinetic term, controlled by the coupling
constant $\gamma$, on the conductivity of the dual field theory.
Along the lines of \cite{Donos:2014cya}, we introduce a perturbation
of the fields of the form
\begin{equation}
ds^{2}=-F\left(  r\right)  dt^{2}+\frac{dr^{2}}{G\left(  r\right)  }%
+r^{2}\left(  dx_1^{2}+dx_2^{2}\right)  +2\epsilon
r^{2}h_{tx_1}\left( r\right) dtdx_1+2\epsilon r^{2}h_{rx_1}\left(
r\right)  drdx_1
\end{equation}
for the metric tensor,
\begin{equation}
A=\mu\left(1-\frac{r_0}{r}\right)dt-\epsilon Edt+\epsilon a_{x_1}\left(  r\right)  dx
\end{equation}
for the gauge field, and
\begin{equation}
\phi_{1}=\mathring{\phi}_{1}+\epsilon\frac{\Phi\left(  r\right)  }{\lambda}%
\end{equation}
for one of the axion fields, with the background axion field fixed
by $\mathring{\phi}_{1}=\lambda x_1$. Here $\mu=Q_e/r_0$ is the chemical potential. Plugging this in the field
equations and keeping the linear terms in $\epsilon$, Maxwell
equations allow one to construct the current density in terms of the
horizon radius $r_h$ as
\begin{equation}
J=\frac{\lambda^{2}r_{h}^{2}+\mu^{2}r_h^2+2^{1-k}\lambda^{2k}r_{h}^{4-2k}\gamma
k}{\lambda^{2}r_{h}^{2}+2^{1-k}\lambda^{2k}r_{h}^{4-2k}\gamma k}E,
\end{equation}
which trivially leads to a DC conductivity of the form
\begin{equation}
\sigma=\frac{\partial J}{\partial
E}=\frac{\lambda^{2}r_{h}^{2}+\mu^{2}r_h^2+2^{1-k}\lambda^{2k}r_{h}^{4-2k}\gamma
k}{\lambda^{2}r_{h}^{2}+2^{1-k}\lambda^{2k}r_{h}^{4-2k}\gamma k}.
\end{equation}
Note that when $\gamma=0$, this expression coincides with the result
obtained in Eq. (4.4) of reference \cite{Donos:2014cya} for minimally coupled axions with standard kinetic terms in $D=4$ (see also \cite{Andrade:2013gsa} and \cite{Jiang:2017imk}). Note that in such a case, in terms of the chemical potential, the DC conductivity remains constant as the radius of the horizon changes. Figure \ref{fig.0} shows the behavior of the DC conductivity for a quadratic derivative self-interaction $(k=2)$ and for a cubic one $(k=3)$. The plots show that in the limit $T\rightarrow 0$ the DC conductivity goes to a constant, which shows that at low temperatures the dual system presents a metallic phase. For large temperatures, the dual system approaches the behavior of the dual system with a minimally coupled axion; i.e. the conductivities saturate to a constant which coincides with the one obtained with $\gamma=0$. As the strength of the nonlinearities of the axions (controlled by $\gamma$) increases, the conductivity at low temperature decreases, and one can see that in such a case larger temperatures are required to recover the result with minimally coupled, free axions.   

\section{Concluding remarks}

As we know, the Horndeski model \cite{Horndeski1} is the most general STT we can construct with second order equations of motion in four dimensions.
In its shift invariant form, it is given by the covariant version of Galileon gravity \cite{Esposito} whose Lagrangian is given by (\ref{lagran}). Black hole solutions with spherical symmetry and with flat asymptotic behavior are forbidden by the no-hair results of Hui and Nicolis \cite{Hui:2012qt}. Their argument relies on the shift invariance of (\ref{lagran}) which forces the scalar field equation of motion to be written as a current conservation law. Then, by demanding that the norm of this current is finite on the horizon of the hypothetical black hole solution it is possible to show that, for spherically symmetric solutions with flat asymptotic geometry, the scalar field must be trivial. In spite of this, for the particular model of the nonminimal coupling between the Einstein tensor and the kinetic term of the scalar field, there are two ways to circumvent the no-hair conjecture. The first one is to relax the asymptotic flatness of the solutions allowing (A)dS behaviors \cite{rinaldi, adolfo2, Minam1}. The second one is to consider scalar fields that do not share the same symmetries of the metric, but are nevertheless nontrivial. In the latter case, the simplest way is to consider scalar fields linearly dependent  on time \cite{charmousis1}. 

In this work we have applied the second strategy in order to construct black brane solutions in the $k$-essence sector of Horndeski/Galileon gravity, specifically in the model in which, along with the standard kinetic term, a nonlinear contribution of the form $X^k$ is included. This sector represents scalar fields with nonlinear kinetic terms without need of any coupling between the scalar field and the curvature.\footnote{This could be interesting due to recent results which indicate that the inclusion of nonminimal couplings with the curvature might induce problems when defining a well-possed initial value problem  in Horndeski theory \cite{Papallo:2017qvl}.}
Specifically speaking, to construct our solutions we have considered
scalar fields that depend linearly on the coordinates of a flat horizon of $(D-2)$-dimensions. The scalar fields are homogeneously
distributed along these flat directions, implying the inclusion of
$i=(D-2)$ scalar fields on the theory. It follows that
each scalar current norm $|J_i|$ does not diverge on the horizon
and, at the same time, satisfies the continuity equation,
$\nabla_{\mu}J_i^{\mu}=0$, with a nontrivial profile for the scalar
field. In order to satisfy the null energy condition and to ensure
the hyperbolicity of the Klein-Gordon equation we have constrained
the possible values of $k$ to be greater than $(D-1)/2$. This is a
sufficient condition to satisfy both requirements. It is interesting
to note that this choice of $k$ endows our solutions with the same
asymptotic behavior of GR without affecting the behavior of the mass term at infinity.
Our solutions possess flat horizon; however the inclusion of the
axion fields provides a new curvature scale including a noncanonical
hyperbolic term on the metric. The electrically and magnetically
charged extension is shown to exist as well as the higher-dimensional extension. We observe that in the case in which
$\Lambda=0$ solutions possessing a cosmological horizon are also
possible provided both mass and  coupling $\gamma$ are negative.
We have analyzed the thermodynamical properties of the asymptotically AdS solutions
in order to study the possible holographic applications and we
have provided an explicit computation for the DC conductivities in
the holographic dual theory of the electrically charged
configurations. It would be also interesting to see whether or not
these solutions violate the reverse isoperimetric inequality
(RII) as is the case of the Horndeski black brane solutions with
axions recently constructed in \cite{Jiang:2017imk} and studied in
\cite{Feng:2017jub}. Another possible extension of this work would be
to see how the nonlinear contribution for the scalar field affects
the realization of the momentum dissipation phenomena extensively
studied for minimally coupled scalar  fields \cite{Andrade:2013gsa}
and also recently in the Einstein coupled sector of Horndeski
gravity \cite{Jiang:2017imk}.

\section{Acknowledgements}
A.C and M.R acknowledge enlightening comments and discussions with Professors. L. Vanzo and S. Zerbini. A.C particularly expresses his gratitude to the Physics Department of the University of Trento for its kind hospitality during the development of this work. A.C.'s work is supported by FONDECYT Grant No. 3150157 and Proyecto Interno UCEN I+D-2016, Grant No. CIP2016. J.O work is funded by CONICYT grant DPI20140053. A.C. and J.O. appreciate the support of the International Center for Theoretical Physics (ICTP), at Trieste, Italy, where part of this work was done.

%%%%%%%%%%%%%%%%%%%%%%%%%%%%%%%%%%%%%%%%%%%%
\section*{Appendix A: D-DIMENSIONAL SOLUTION}
%%%%%%%%%%%%%%%%%%%%%%%%%%%%%%%%%%%%%%%%%%%%
The higher-dimensional neutral extension of the solution is given by
\begin{equation}
ds^2=-F(r)dt^2+\frac{dr^2}{F(r)}+r^2d\Sigma^2_{D-2}
\end{equation}
where $d\Sigma_{D-2}$ stands for a (D-2) base manifold with null
curvature, and where 
\begin{equation}
F(r)=\frac{r^2}{l^2}-\frac{2M}{r^{D-3}}-\frac{\lambda^2}{2(D-3)\kappa}-
\gamma\frac{\lambda^{2k}}{2^k(2k+1-D)}r^{2(1-k)}.
\end{equation}
Here, the axion fields are $\phi_i=\lambda x_i$ and
$l^{-2}:=-\frac{2\Lambda}{(D-2)(D-1)}$. As it was also clear in four
dimensions with $k=3/2$, there exists a logarithmic branch for
$k=(D-1)/2$.\footnote{As we state on the four dimensional case please note that these solutions have a standard AdS asymptotic for $k>(d-1)/2$.} In four dimensions, the logarithmic branch reads
\begin{equation}
F(r)=\frac{r^2}{l^2}-\frac{2M}{r}-\frac{\lambda^2}{2\kappa}-\gamma\frac{\sqrt{2}}{4}\frac{\lambda^3
\ln(r/r_0)}{\kappa r}.
\end{equation}
It is easy to see that these solutions can be easily
extended to the charged cases.

\end{document}